\documentclass{aa}
\usepackage{graphics,psfig}
\begin{document}

   \thesaurus{11    
              (11.05.2; 
               11.06.1; 
               11.06.2; 
               11.16.1; 
               11.19.6; 
               11.19.2
              )} 
   \title{Cut-off radii of galactic disks
}
\subtitle{A new statistical study on the truncation of galactic disks
\thanks{
Based on observations collected at the European Southern Observatory, Chile 
and Lowell Observatory, Flagstaff (AZ), USA
}
}
\author{M. Pohlen \and R.-J. Dettmar \and R. L\"utticke}
\institute{Astronomisches Institut, Ruhr-Universit\"at Bochum, 
           D-44780 Bochum, Germany\\
              email: (pohlen,dettmar,luett)@astro.ruhr-uni-bochum.de}
   \date{Received 13 July 1999 / Accepted 22 March 2000}
\maketitle

   \begin{abstract}
We present the analysis of a CCD survey of 31 nearby ($\le$ 110 Mpc) 
edge-on spiral galaxies. The three-dimensional one-component best 
fit models provide their disk-scalelengths $h$ and for the first 
time their disk cut-off radii $R_{co}$. 
We confirm for this sample the existence of such sharp truncations, and find 
a significantly lower mean value of the distance independent 
ratio $R_{co}/h =2.9 \pm 0.7$ than the standard value of $4.5$ often used in 
the literature. 
Our data show no correlation of these parameters with Hubble type, whereas we 
report a correlation between $R_{co}/h$ and the distance based scalelength in 
linear units.
Compared to the Milky Way we find only lower values of $R_{co}/h$,
explained either by possible selection effects or by the completely different 
techniques used.   
We discuss our data in respect to present models for the origin of the cut-off
radii, either as a relict of the galaxy formation process, or as an 
evolutionary phenomenon. 
   \end{abstract}
\section{Introduction}
Although {\it cut-off radii} of spiral galaxies are known for about 20 years
no unique physical explanation has been given to describe this observational 
phenomenon. They were already mentioned by van der Kruit (\cite{vdk79}), who 
stated, based on photographic material, that the outer parts of disks of 
spiral galaxies ''do not retain their exponential light distribution to such 
faint levels'', whereas the exponential behaviour of the radial light 
distribution for the inner part was well accepted 
(de Vaucouleurs \cite{devau59}, Freeman \cite{free}).
For three nearby edge-on galaxies he claimed, that the typical radial 
scalelength $h$ steepens from 5 kpc to about 1.6 kpc at the edge of the disk. 
This is confirmed by modern deep CCD imaging 
(Abe et al. \cite{abe}, Fry et al. \cite{fry}, 
N\"aslund \& J\"ors\"ater \cite{naes}). 
In a fundamental series of papers van der Kruit \& Searle
(\cite{vdk81a}, \cite{vdk81b}, \cite{vdk82a}, \cite{vdk82b}) determined a 
three dimensional model for the luminosity density of the old disk 
po\-pu\-lation taking into account these sharp truncations at the 
{\it cut-off radius} $R_{co}$.
They applied their model of a locally isothermal, selfgravitating, and 
truncated exponential disk to a sample of seven edge-on galaxies and found 
that all disks show a relatively sharp cut-off where the scalelength $h$ 
suddenly drops below 1 kpc, starting at radii of $(4.2 \pm 0.5) h$.
The {\it cut-off radius} of edge-on galaxies is detected at levels of 
24-25 mag$/\sq\arcsec$ which is about 2-3 mag brighter compared to face-on 
disks due to the integration along the line of sight.
Therefore van der Kruit \& Shostak (\cite{vdkshos82}) and Shostak \& van der 
Kruit (\cite{shos}) quote the only known cut-offs in the literature for 
face-on galaxies. In addition to the much lower brightness one has to deal 
with intrinsic deviations from the circular symmetry of the disk, for example 
from the young stellar population, hidden by an azimuthally averaged profile. 
In a subsequent paper van der Kruit (\cite{vdk88}) stated that out of the 20 
face-on galaxies observed by Wevers et al. (\cite{wev}) only four did not 
show any sign for a drop off as judged from the last three contours.
Barteldrees \& Dettmar (\cite{bdold}) confirmed for the first time the 
existence of these truncations for a larger sample of edge-on galaxies using 
CCD surface photometry refining the previous photographic measurements.\\
These truncations are not the boundary of the galactic baryonic mass 
distribution, but such 'optical edges' suggest dynamical consequences for the
interpretation of observed rotation curves (Casertano \cite{caser}), as 
well as for the explanation of warped disks (Bottema \cite{bott}).  
Their sharpness restrict the radial velocity dispersion at the edge of the disk
(van der Kruit \& Searle \cite{vdk81a}), and will therefore have important 
implication for viscous disk evolution (Thon \& Meusinger \cite{thon}).
According to Zhang \& Wyse (\cite{zhang}) the disk cut-off radii constrain
the specific angular momentum in a viscous galaxy evolution scenario.\\
In this letter we report the largest sample of well defined cut-off radii for 
edge-on galaxies derived by CCD surface photometry. 
Our sample (Pohlen et al. \cite{pohl}, Paper II) comprises 31 galaxies, 
including the 17 galaxies of Barteldrees \& Dettmar (\cite{bd}, hereafter 
Paper I).  
Thereby we are able to derive first statistical conclusions  
and determine general correlations with other characteristic galaxy parameters 
in order to approach in the future a physical model explaining the observed 
phenomenon.
%
\section{Observations and reduction}
Sample selection, observations, and data reduction are described in detail
in Paper I and II, and will be repeated only briefly here.
We have compiled a homogeneous set of 31 galaxies with well defined models of 
their three-dimensional disk luminosity distribution, out of our sample of 
about 60 highly inclined disk galaxies, selected from the UGC 
(Nilson \cite{ugc}) and ESO-Lauberts \& Valentijn (\cite{lv}) catalog. 
The data were obtained at the ESO/La Silla 2.2m and the Lowell Observatory 
42-inch telescope. Images are taken either in Gunn g, r, i, or Johnson R 
filters, with resulting pixel scales of 0.36$''$ and 0.7$''$, respectively.
After standard reduction, images were rotated to the major axes of the disk. 
Although most of the images were taken during non-photometric nights we have 
tried to perform photometric calibration for each image by comparing 
simulated aperture measurements with published integrated aperture data, 
resulting in the best possible homogeneous calibration for the whole sample. 
The resulting typical values for the limiting surface brightness measured by a 
three sigma deviation on the background are: 
$\mu_{g}\!\approx\!25$mag$/\sq\arcsec$, 
$\mu_{R,r}\!\approx\!24.5$mag$/\sq\arcsec$, and 
$\mu_{i}\!\approx\!23.5$mag$/\sq\arcsec$.
In order to obtain absolute values of the determined structural parameters,
we estimated the distance of our galaxies according to the Hubble relation
($H_{0}\!=\!75$ km s$^{-1}$Mpc$^{-1}$) using published heliocentric radial 
velocities corrected for the Virgo centric infall.
\section{Analysis}
\subsection{Disk model and fitting}
Our model, as described in detail in Paper I and II, for the three-dimensional
luminosity distribution for galactic disks is based upon the fundamental work 
of van der Kruit and Searle (\cite{vdk81a}): 
\begin{equation}
\hat{L}(R,z) = \hat{L}_0 \ \exp{\left(-\frac{R}{h}\right)} \ f_n(z,z_0) 
\ {\rm H}(R_{co}-R)
\label{hatl}
\end{equation}
$\hat{L}_0$ is the central luminosity density in units of 
[$L_{\sun}$ pc$^{-3}$], $R$ and $z$ are the radial resp. vertical axes in 
cylindrical coordinates, $h$ is the radial scalelength, and $z_0$ the 
scaleheight. 
$f_n(z,z_0)$ describes three different fitting functions for the vertical
distribution: exponential, sech, and the physically motivated isothermal 
(sech$^2$) case (van der Kruit \cite{vdk88}). 
$R_{co}$ is the cut-off radius, where the stellar luminosity density is 
assumed to be zero outside, mathematically expressed by a Heaviside function 
H$(x_0-x)$.  
These radii are defined at the position where the radial profiles bend nearly 
vertical into the noise, whereby a mean value is taken for the two different 
sides.\\
Depending on the inclination angle $i$, we numerically integrate this 3D-model 
along the line of sight and compare the two-dimensional result with the 
observed CCD image, leading to six free fitting parameters:
the inclination $i$, the central luminosity density $\hat{L}_0$, the 
scalelength $h$, and -height $z_0$, the cut-off radius $R_{co}$, and the 
function for the z-distribution $f_n(z,z_0)$.
After our discussion about two different fitting methods in Paper II, we 
finally used our implemented ``down\-hill simplex-method'' 
(Nelder \& Mead, \cite{nm}) to minimize the difference between model and 
observed disk. 
The possible influence of these parameters on the neglected dust distribution 
is estimated in Paper II.
\section{Results}
\subsection{Distribution of cut-off radii}
\begin{figure}[t]
\psfig{figure=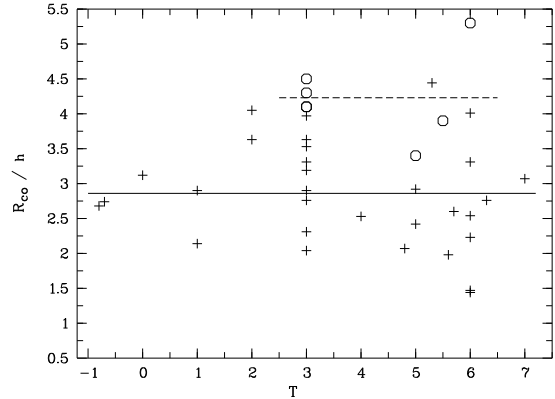,height=5.5cm,angle=270}
\caption{Ratio of cut-off radius to radial scalelength versus morphological 
type T. The 31 galaxies from our sample are marked by a cross, whereas the 
circles are the values for 7 galaxies from van der Kruit \& Searle 
(\cite{vdk82a}). The dashed line represents their and the solid line our mean 
value.}
\label{rcodhvert}
\end{figure}
\begin{figure}[t]
\psfig{figure=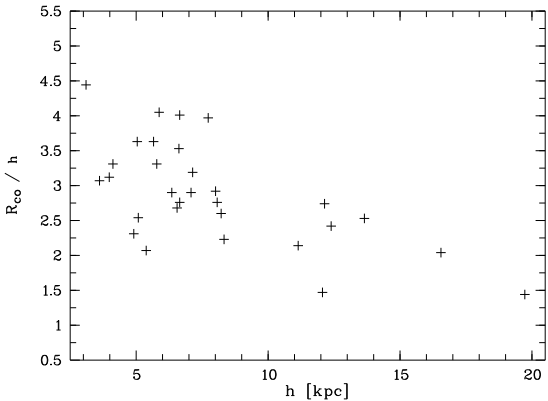,height=5.5cm,angle=270}
\caption{Ratio of cut-off radius to radial scalelength versus 
scalelength in absolute units for the 30 galaxies with measured radial 
velocities.}
\label{rcodhverhkpc}
\end{figure}

Understanding the phenomenon of cut-off radii in galactic disk requires 
as an essential step a statistical study of galaxies covering the Hubble 
sequence. 
Figure \ref{rcodhvert} shows, already suggested by Barteldrees \& Dettmar 
(\cite{bdold}) for a smaller sample of 20 galaxies, that the distance 
independent ratio of cut-off radius to radial scalelength is significantly 
lower than derived from the often referred sample of van der Kruit \& Searle 
(\cite{vdk82a}) with 7 galaxies. They reported a mean value of 
$R_{co}/h=4.2\pm0.5$ (ranging from 3.4 to 5.3), whereas our sample gives a 
ratio of $R_{co}/h=2.9\pm0.7$ (1.4$-$4.4) even below their minimal value. 
As obvious from Fig.\,\ref{rcodhvert} this difference is not caused by the 
larger range of Hubble types covered by our sample. 
The estimated error for this ratio due to the selection of the best fitting 
model described in Paper II is $\pm 0.6$  and has the same order as 
the quoted standard deviation. 
As shown in Paper II for two different dust distributions with values 
observed by Xilouris et al. (\cite{xil99}), the influence of the 
neglected dust on our fitting process will be an overestimation of the 
scalelength $h$, whereas $R_{co}$ is independent. For the {\it worst} case, 
defined by the largest measured values for $h_d/h_{\rm *}$, $z_d/z_{\rm *}$, 
and $\tau_R$, we find that this will chance our values for $R_{co}/h$ by 
$+0.5$. Applied to the mean we are in this case still $0.8$ below the value
of van der Kruit \& Searle (\cite{vdk82a}).
We do not find a correlation between $R_{co}/h$ and the Hubble type,
although it should be mentioned that in general for galaxies later than Scd
the fitting process is strongly affected by intrinsic variation, e.g. 
individual bright HII-regions, which makes it impossible to fit our simple 
symmetric model. On the other side some early type galaxies and particularly 
lenticulars do not show any evidence for a cut-off at all. This is already 
suggested by van der Kruit (\cite{vdk88}) observing some early type face-on 
galaxies and will be discussed in detail in a forthcoming paper. \\ 
Figure \ref{rcodhverhkpc} shows a possible correlation between $R_{co}/h$ and 
the scalelength in absolute units: 
{\it Large disks with regard to their scalelengths $h$ are short in terms of their cut-off radii.} 
Together with the fact that the cut-off occurs, within the errors of 
$\pm 0.5$mag, at nearly the same surface brightness level, this can be 
explained with a correlation between the central surface brightness and the 
scalelength of the galaxy, recently proposed by Scorza \& van den Bosch 
(\cite{scovdb}) for galactic disks of different sizes. 
\subsection{Comparison with literature}    
For the 30 galaxies with known radial velocities we find 
values for the scalelength of 3.1 up to 19.7 kpc with a median of 6.6 kpc. 
Van der Kruit (\cite{vdk87}) determines for a diameter limited sample
of 51 galaxies scalelengths in the range of $0.7 - 9.2$ kpc with a maximum of 
the distribution at about 3 kpc. De Jong (\cite{dejong}) derives for his 
sample of 86 face-on galaxies transformed to our $H_{0}$ a range of 
$1.0-14.4$ kpc with a median around $3.0$ kpc, whereas Courteau 
(\cite{cour}) finds for 290 Sb-Sc galaxies a range of $0.5 - 9.6$ kpc with a 
maximum at 3.9 kpc; reduced to our $H_{0}$. In agreement with de Jong 
(\cite{dejong}) we do not find a correlation of the scalelength with the 
Hubble type. \\
We find cut-off radii from $11.1 - 34.5$ kpc with 
a median at 20.2 kpc, compared to the only available sample of cut-off radii 
by van der Kruit \& Searle (\cite{vdk82a}) with $7.8 - 24.9$ kpc for their 7 
investigated galaxies. Although we do not find a tight correlation between
catalogued surface brightness radii, e.g. $D_{25}$, and our cut-off radii, they
can be used to compare the sizes of the galaxies within our sample.
Rubin et al. (\cite{rubin}) study 21 Sc galaxies, where they claim radii,
characterized by the radius at the $D_{25}$ contour reduced to our $H_{0}$,
of 81.3 kpc and 35.3 kpc for the two biggest ones, and Romanishin 
(\cite{roma}) finds values of $30-73$ kpc for 107 intrinsically large spiral 
galaxies. \\
We find a clear correlation between the determined cut-off radius and the 
distance of the galaxy. This implies that we pick intrinsically large galaxies 
at higher distances due to our selection criterion which is based on the 
angular diameter matching the filed of view.
\subsection{Comparison with the Milky Way}
It is of particular interest to compare our statistical result with the 
structural parameters derived for the Milky Way. 
Robin et al. (\cite{robin}) as well as Ruphy et al. (\cite{ruphy}) determine 
the radial structure of the galactic disk with a synthetic stellar population 
model using optical and NIR star-counts, respectively. 
They confirm a sharp truncation of the old stellar disk at $14\pm0.5$ kpc 
and $15\pm2$ kpc, respectively. Freudenreich (\cite{freu}) fits a model for 
the old galactic disk to the NIR data obtained from the survey of the DIRBE 
experiment also confirms an outer truncation of the 
disk around $12.4 \pm 0.1$ kpc. The result of both methods depend directly on 
the distance to the galactic center ($R_{0}=8.5 \pm1$ kpc).
These values are in agreement with the findings of Heyer et al. 
(\cite {heyer}), who measure a sharp decline in the CO mass surface density 
and conclude that the molecular disk is effectively truncated at 
$R\!=\!13.5$ kpc. \\
In contrast to former investigations (van der Kruit \cite{vdk86}, 
Lewis \& Freeman \cite{lewis}, Nikolaev \& Weinberg \cite{nik}) placing the 
Milky Way scalelength around $4-5.5$ kpc, Robin et al. (\cite{robin}), 
Ruphy et al. (\cite{ruphy}), and Freudenreich (\cite{freu}) quote 
significantly lower scalelengths of $2.5\pm 0.3$ kpc, $2.3\pm0.1$ kpc, and 
$2.59 \pm 0.02$ kpc, respectively. 
This leads to values of $5.6\pm0.5$, $6.5\pm1.2$, and $4.8\pm0.1$ for 
$R_{co}/h$. Whereas the first two values are significantly higher than any 
value found in our sample (even the highest value of van der Kruit \& Searle 
is only 5.3) the latter determination by Freudenreich is consistent with our 
highest value of 4.4 within the errors. \\
If the Milky Way is a 'typical' galaxy with  $R_{co}/h=2.9$ the scalelength 
should be expected to be $h \ge 4.1$ kpc for $R_{co}\ge 12$ kpc. 
\subsection{Comparison with models}
Only few theoretical models can be found in the literature addressing  
a physical description for the origin of cut-off radii. \\
Taking into account a basic picture of galaxy formation, starting 
with a rotating protocloud, Seiden et al. (\cite{seiden}) explain in their 
framework of a stochastic, self-propagating star-formation theory (SSPSF)
several properties of galactic disks. The crucial point is, that they 
assume a $1/R$ dependence instead of an exponential law for the total surface 
density. In this case they show that a feature similar to a cut-off radius 
automatically appears in the radial profile, which is directly linked with the 
scalelength. This is in contrast to Fig.\,\ref{rcodhverhkpc}, 
where $R_{co}$ and $h$ vary independently. \\  
Van der Kruit \& Searle (\cite{vdk81a}) proposed that within a scenario of 
slow disk formation (Larson \cite{larson}) this radius might be that radius 
where disk formation time equals the present age of the galaxy. 
This isolated slow evolution is in contradiction to recent models preferring 
interaction and merging as a driver for galaxy evolution 
(Barnes \cite{barnes}). \\
Later van der Kruit (\cite{vdk87}) proposed a working hypotheses which 
already includes some of the currently accepted ingredients for galaxy 
formation to explain the truncation as a result of the formation process.
Galactic disks develop from collapsing, rotating proto-clouds.
After the dark matter has settled into an isothermal sphere first 
star-formation in the center builds up a bulge component and the remaining
material settles in gaseous form with dissipation in a flat disk under 
conservation of specific angular momentum. This leads to a constant value
for $R_{co}/h$ of 4.5, which is in contrast to our observations. \\
In a recent paper about galaxy formation and viscous evolution Zhang \& 
Wyse (\cite{zhang}) additionally consider a self-consistent description of
the disk-halo system by dropping the assumption of a static halo and find
that the disk cut-off radii indeed constrain the specific angular momentum. \\
Kennicutt (\cite{kenni}) shows that for a sample of 15 face-on spiral 
galaxies, analysing HI, CO and H$_{\alpha}$ data, star-formation stops below
a critical threshold value, which is associated with large scale 
gravitational instabilities. Taking into account the dynamical critical gas 
density $\Sigma_{{\rm crit}}$ for a thin, rotating, isothermal gas disk 
proposed by Toomre (\cite{toom}) he observes the abrupt decrease in 
star-formation at a radius where the measured gas density drops below 
$\Sigma_{{\rm crit}}$. In the case of NGC 628 this radius coincides with 
$R_{co}$ determined by Shostak \& van der Kruit (\cite{shos}).\\
Although it is still unknown if the cut-off radius is an evolutionary 
phenomenon or has its origin in the galaxy formation process a star-formation 
threshold at the 'optical edge' seems to be a promising approach to address 
this problem (Elmegreen \& Parravano \cite{elm}, and references therein;
Ferguson et al. \cite{ferg}). 
This will be done in the future by enlarging the sample with a better defined  
selection criterion which also includes the environment. 
\end{document}